# NEW METHOD OF DETERMINATION OF THE TILT ANGLE AT A LIQUID CRYSTAL – GLASS INTERFACE

**A. A. Karetnikov, N. A. Karetnikov, A. P. Kovshik, Y. I. Rjumtsev**

*Department of Physics, St. Petersburg State University, St. Petersburg, Petrodvorets, 198504, Russia*

E-mail: akaret@mail.ru

A polarimetric method is considered for measuring the tilt angle of the LC director at an LC layer interface. The method involves the use of an LC cell operating in the mode of the total internal reflection of the ordinary wave. The method is based on measuring the angle of the plane of light incidence on the cell with the polarization vector of the extraordinary wave passed through the cell or with the polarization vector of the ordinary wave reflected from the cell. To calculate the tilt angle, the polarization azimuth of the light incident on the cell, which induces only the ordinary wave, can be also used. The method is applicable throughout the whole range of the LC director tilt angles. The data on the LC director tilt angles measured in several cells are presented.

Key words: Liquid crystal, tilt angle, polarization azimuth.

## Introduction

The optical properties of various LC devices essentially depend on the orientation of the LC director at the layer interface. Therefore, it is important to provide an option for convenient and fast determination of the director tilt angle.

Various methods of determination of the tilt angle were presented by Scheffer and Nehring [1]. The cell rotation method is conventionally used for LC layers uniformly oriented within the 0°- 20° and 70°- 90° ranges [2]. This method calls for the layer thickness and LC refractive indices to be known. An extended version of this method that enables determination of the tilt angle within the 0º - 90º range has been recently developed [3]. The null-magnetic method is very sensitive and independent of the LC constants; however, it involves the use of a large magnet [1]. Both the cell rotation method and the null-magnetic method are applicable to the cells with a 90-degree twisting [1, 2, 4, 5]. The tilt angle can be calculated from the critical angle of the total internal reflection of the extraordinary wave provided that the values of the refractive indices are known [6-7]. A modern conoscopic method can be used for a uniformly oriented LC layer [8].

The polarimetric method that we developed was briefly described in [9]. This paper presents a refined and complemented description of the proposed method. The method is based on measuring the angle of the plane of light incidence on the LC cell with the polarization vector of the extraordinary wave passed through the LC cell or with the polarization vector of the ordinary wave reflected from the cell due to the total internal reflection. In contrast to the conventional cell rotation method, in our method, the LC director tilt angles at the opposite LC – glass interfaces in the cell may be different, but at the both interfaces the LC director should be located in a same plane perpendicular to the light incidence plane.

**Method**

In the method considered here, the LC cell consists of two trapezoidal glass prisms with an LC layer embedded between them. With the specially selected LC - glass refractive index ratio, this cell transmits the extraordinary wave and reflects the ordinary wave due to the total internal reflection effect [10-13]. The position of the planes of polarization of these waves depends of the orientation of the LC director at the LC—glass interfaces. In our method, the measured quantity is the polarization azimuth, that is, the angle of the plane of light incidence with the polarization vector of the light wave (the electric field vector of the light wave), which is the ordinary or extraordinary wave entering or leaving the cell. If the director at both LC interfaces is located in the plane normal to the plane of light incidence, the director tilt angle can be uniquely determined using the values of the polarization azimuths of these waves.

The schematic layout of the method in the Cartesian coordinate system is depicted in Fig. 1.

It is assumed that the LC layer of thickness d is located between the plane surfaces $P_1$ and $P_2$ of two semi-infinite glasses. The LC directors $\mathbf{n_1}$ and $\mathbf{n_2}$ at the surfaces $P_1$ and $P_2$ lie in the YOZ plane at angles $\psi_1$ and $\psi_2$ with normal $\boldsymbol{N}$, respectively

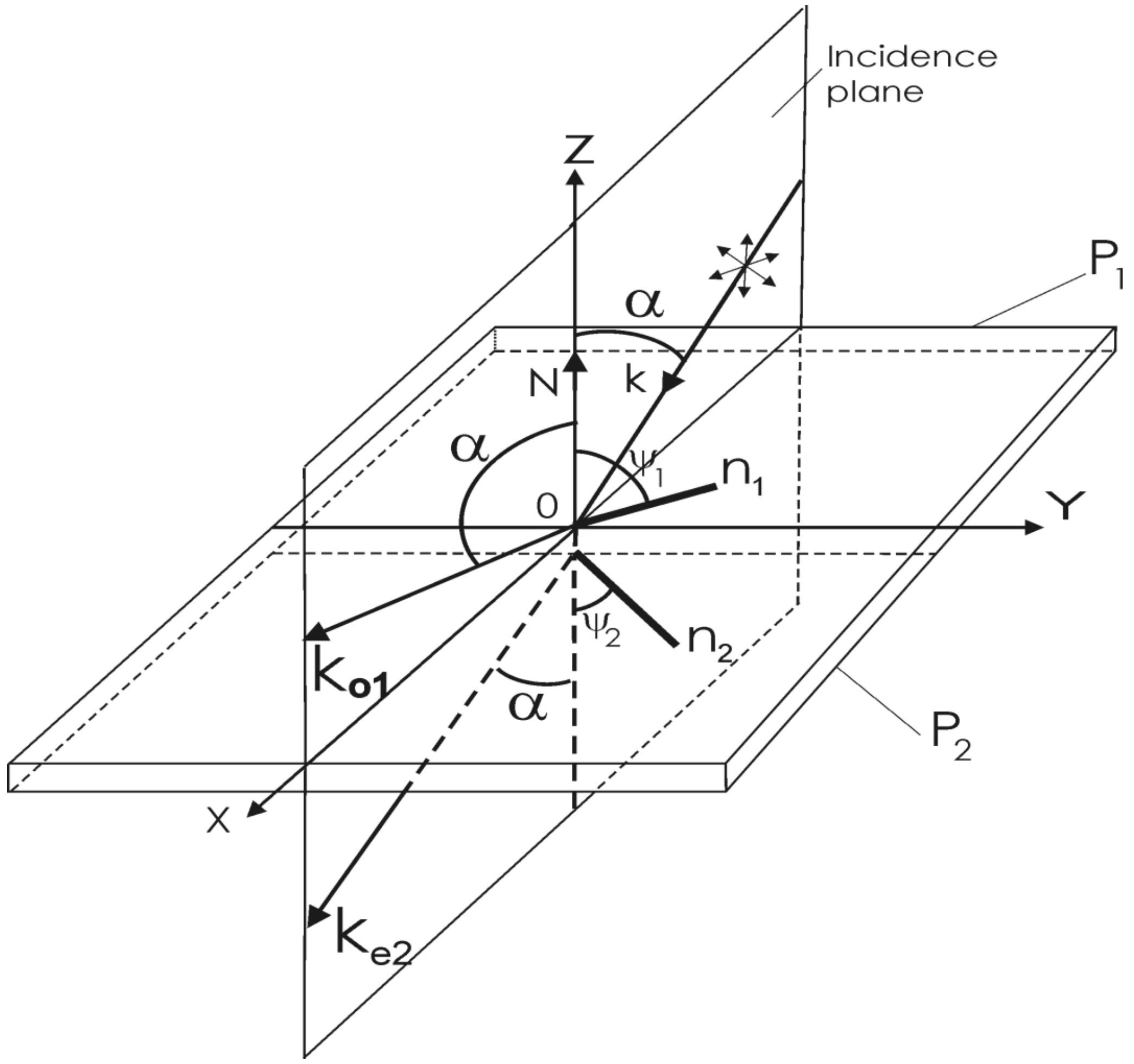

FIGURE 1. Geometrical schematic of the method.

Planes $P_1$ and $P_2$ are the glass—LC interfaces; $\mathbf{n_1}$ and $\mathbf{n_2}$ are the LC directors at the LC layer interfaces (vectors $\mathbf{n_1}$ and $\mathbf{n_2}$ are shown outside the LC layer); **k** is the wave vector of the incident wave; $\alpha$ is the incidence angle; $\mathbf{k_{o1}}$ and $\mathbf{k_{e2}}$ are the wave vectors of the ordinary wave (reflected from surface $P_1$) and the extraordinary wave (passed through surface $P_2$), respectively; and angles $\beta_{o1}$ and $\beta_{e2}$ are the polarization azimuths of these waves.

It is assumed that the LC layer of thickness d is located between the plane surfaces $P_1$ and $P_2$ of two semi-infinite glasses. The LC directors $\mathbf{n_1}$ and $\mathbf{n_2}$ at the surfaces $P_1$ and $P_2$ lie in the YOZ plane at angles $\psi_1$ and $\psi_2$ with normal $\boldsymbol{N}$, respectively ($0\leq\psi_1, \psi_2\leq\pi/2$ ). The YOZ plane is normal to the plane of light incidence. Let an unpolarized plane light wave (with wave vector **k**) be incident on the surface $P_1$ at angle α with normal **N** in the XOZ plane. The refractive indices and angle of incidence α are chosen so that the following relationship is satisfied:

$$\frac{N_o}{N_{gl}} < \sin\alpha < \frac{N_e}{N_{gl}} \qquad (1)$$

where $N_o$ and $N_e$ are the principal LC refractive indices, and $N_{gl}$ is the glass refractive index. In turn, the birefringence $\Delta N = N_e - N_o$ and thickness d of the LC layer complies with the Mogen relationship:

$$\frac{\Delta N \cdot d}{\lambda} >> \frac{|\Psi_1 - \Psi_2|}{\pi} \qquad (2)$$

Condition (1) means that the ordinary wave is reflected due to the total internal reflection at the $P_1$ glass—LC interface (wave vector $\mathbf{k_{o1}}$), whereas the extraordinary wave passes through the LC layer and enters into the glass through the $P_2$ interface (wave vector $\mathbf{k_{e2}}$). Condition (2) means that the linear polarization of the extraordinary wave is retained in the LC layer and the polarization vector of this wave is rotated according to a change in the director orientation.

For the extraordinary wave passed through the LC layer and entered into the glass (Figure 2),

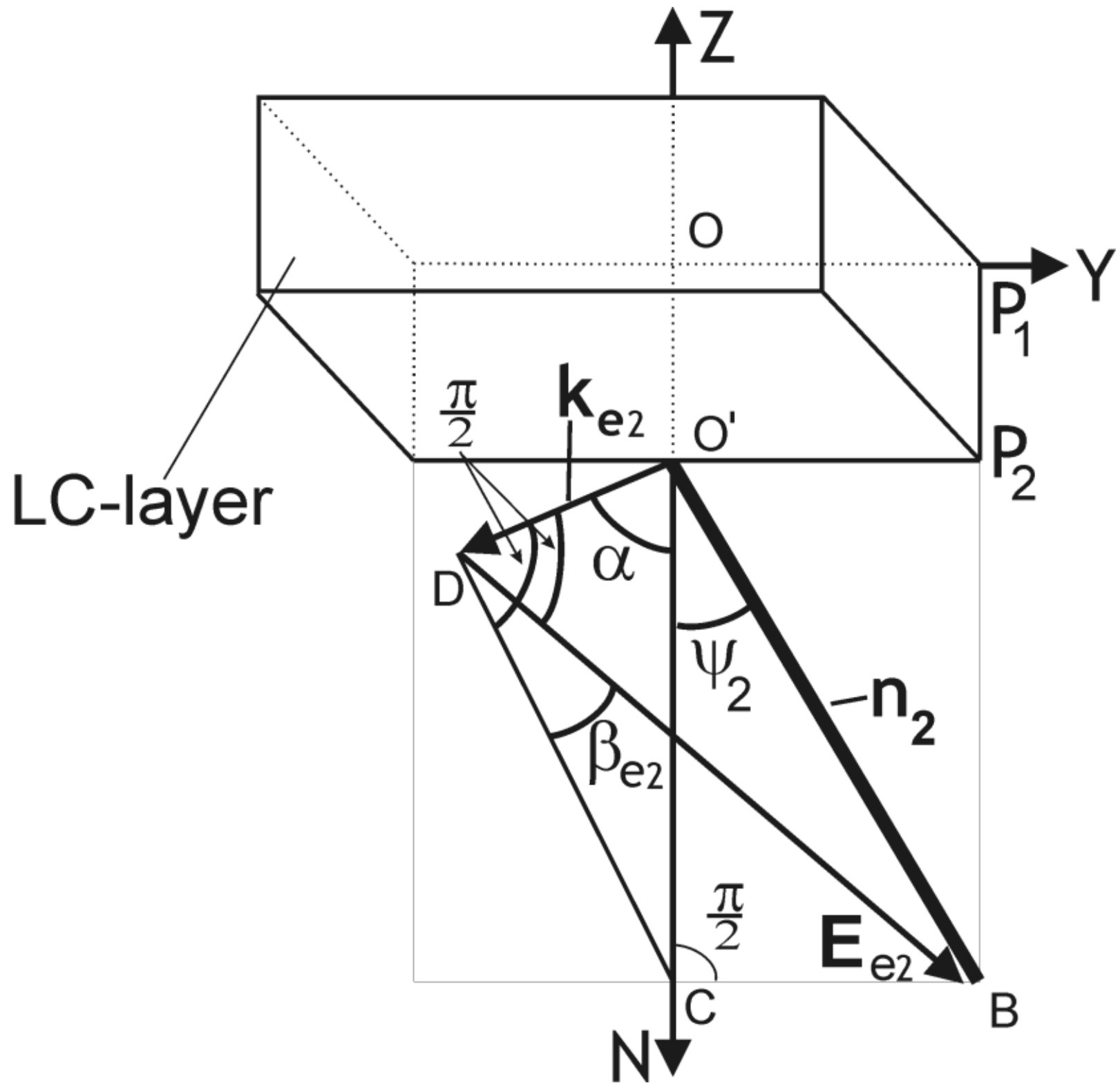

FIGURE 2. Schematic representation of the geometric relationship between angles α, $\beta_{e2}$ , and $\Psi_2$ for the extraordinary wave passed through surface $P_2$.

the electric field vector $\mathbf{E_{e2}}$ is normal to the wave vector $\mathbf{k_{e2}}$ and lies in the plane containing $\mathbf{k_{e2}}$ and director $\mathbf{n_2}$ [14, 15]. Vector $\mathbf{E_{e2}}$ makes azimuth angle $\beta_{e2}$ with the plane of incidence (CD is the $\mathbf{E_{e2}}$ component on the plane of incidence). Using the right triangles BCO′, CDO′, and BCD formed by vectors $\mathbf{k_{e2}}$, $\mathbf{n_2}$, **N**, and $\mathbf{E_{e2}}$, we can find the relationship between the measured angle $\beta_{e2}$ and the tilt angle $\psi_2$:

$$\mathrm{tg}\Psi_2 = \mathrm{tg}\beta_{e2} \cdot \sin\alpha \qquad (3)$$

If the unpolarized light wave is incident from glass on the opposite side of the LC layer, namely, on the $P_2$ interface, at the same angle and in the same plane ZOX and emerges through the $P_1$ interface, we similarly obtain:

$$\mathrm{tg}\Psi_1 = \mathrm{tg}\beta_{e1} \cdot \sin\alpha \qquad (4)$$

where $\beta_{e1}$ is the azimuth angle of the polarization vector $\mathbf{E_{e1}}$ of the extraordinary wave emerged from the LC layer into glass.

Let us consider the ordinary wave reflected from the surface $P_1$. Its wave vector $\mathbf{k_{o1}}$ coincides in direction with the wave vector $\mathbf{k_{e1}}$ of the extraordinary wave emerged from the LC layer through the surface $P_1$ (Figure 3). The polarization vector $\mathbf{E_{o1}}$ of the ordinary wave makes angle $\beta_{o1}$ with the plane of incidence.

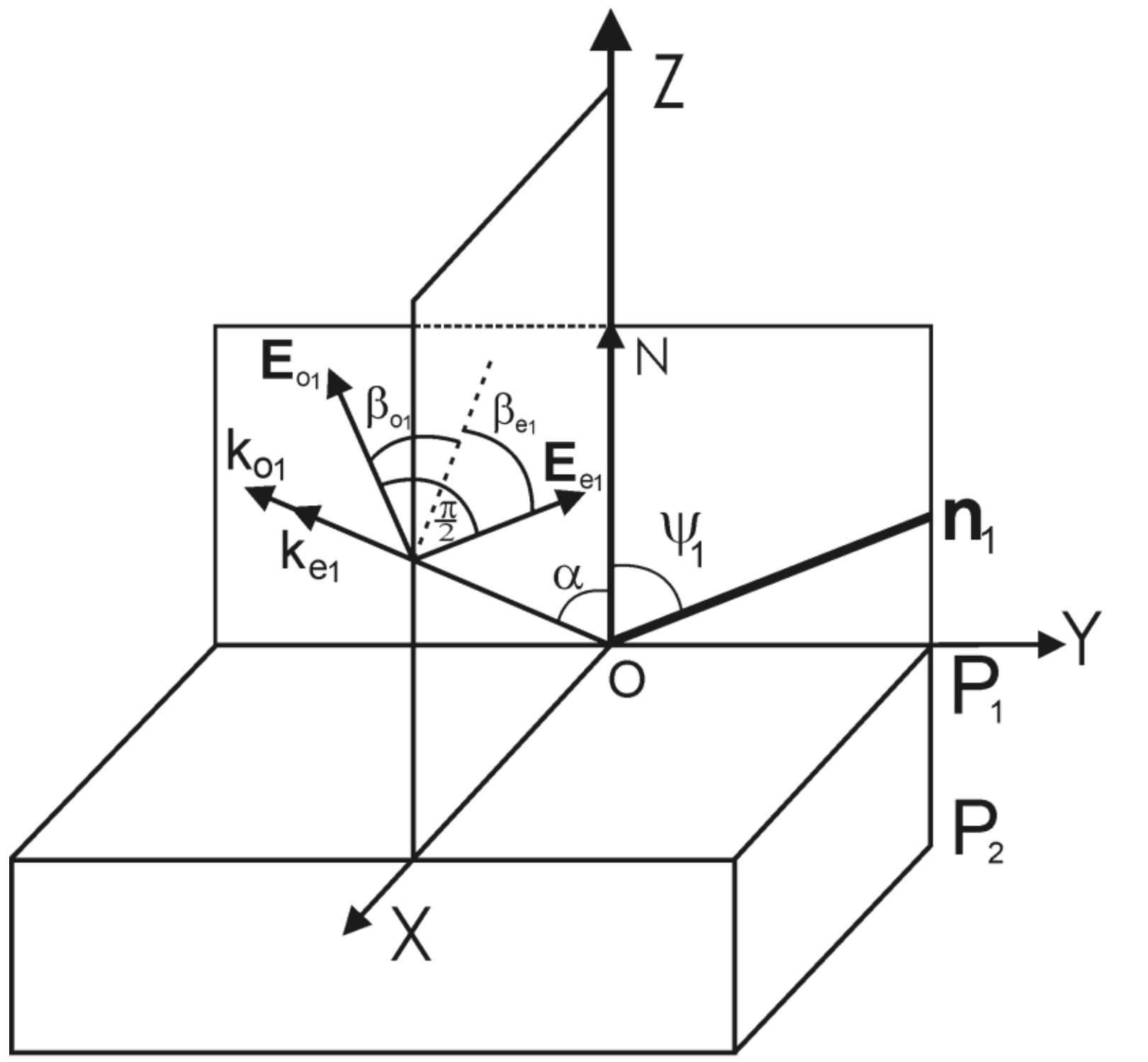

FIGURE 3. Schematic representation of the geometric relationship between angles $\alpha$, $\beta_{o1}$, and $\Psi_1$ for the ordinary wave reflected from surface $P_1$

Vector $\mathbf{E_{o1}}$ is orthogonal to the wave vector $\mathbf{k_{o1}}$, director $\mathbf{n_1}$ [14] and wave vector $\mathbf{k_{e1}}$. Hence, it naturally follows that vector $\mathbf{E_{o1}}$ is orthogonal to vector $\mathbf{E_{e1}}$:

$$\beta_{o1} + \beta_{e1} = \frac{\pi}{2} \tag{5}$$

Same relationship appears for azimuth $\beta_{o2}$ of the ordinary wave reflected from the surface $P_2$:

$$\beta_{o2} + \beta_{e2} = \frac{\pi}{2} \tag{6}$$

By substituting $\beta_{e1}$ and $\beta_{e2}$ from Eqs. (5) and (6) in Eqs. (4) and (3), we obtain the following equations relating the tilt angles $\psi_1$ and $\psi_2$ to the polarization azimuths $\beta_{o1}$ and $\beta_{o2}$:

$$\mathrm{tg}\Psi_1 = \mathrm{ctg}\beta_{o1} \cdot \sin\alpha \tag{7}$$

$$\mathrm{tg}\Psi_2 = \mathrm{ctg}\beta_{o2} \cdot \sin\alpha \tag{8}$$

Thus, equations (3)-(4) and (7)-(8), which relate the director tilt angle to corresponding polarization azimuths, have simple form and do not include refractive indices and layer thickness.

However, the knowledge of the refractive indices of a liquid crystal and glass may be needed if the extraordinary wave is used to determine the director tilt angle. This is associated with the fact that the interface between the two media is characterized by different transmittances for the components of electric field vector lying in the plane of incidence and normal to this plane. The difference in the transmittances for the vector components results in the measured value of the polarization azimuth different from its true value. When the refractive indices are known, the measured azimuth of the extraordinary wave can be recalculated to the true azimuth using Fresnel equations.

The presence of the Fresnel reflections should be also taken into account when the director tilt angles are determined using the ordinary wave. If the incident light is unpolarized, the extraordinary wave partially reflected on two anisotropic interfaces is added to the totally reflected ordinary wave. Therefore, the measured polarization azimuth of the reflected light differs from that of the ordinary wave. It is possible to avoid Fresnel reflections of the extraordinary wave using linearly polarized incident light with such a polarization azimuth that induces **only** the reflected ordinary wave in the LC layer. This reflected ordinary wave is elliptically polarized [14, 15]. For the correct measurement of the wave azimuth $\beta_{o1}$ (or $\beta_{o2}$) it is necessary to compensate its ellipticity (for example, by a quarter-wave plate), or measure the polarization azimuth of the **incident** wave instead of that of the reflected wave. In the latter case, the value of the polarization azimuth is also substituted in equation (7) or (8). Therefore, the knowledge of the LC refractive indices is not needed for determination of the LC director tilt angle by the azimuths of the reflected or incident waves.

## Experiment

The developed method was applied to measuring the director tilt angle in several samples.

The experimental setup is schematically depicted in Figure 4.

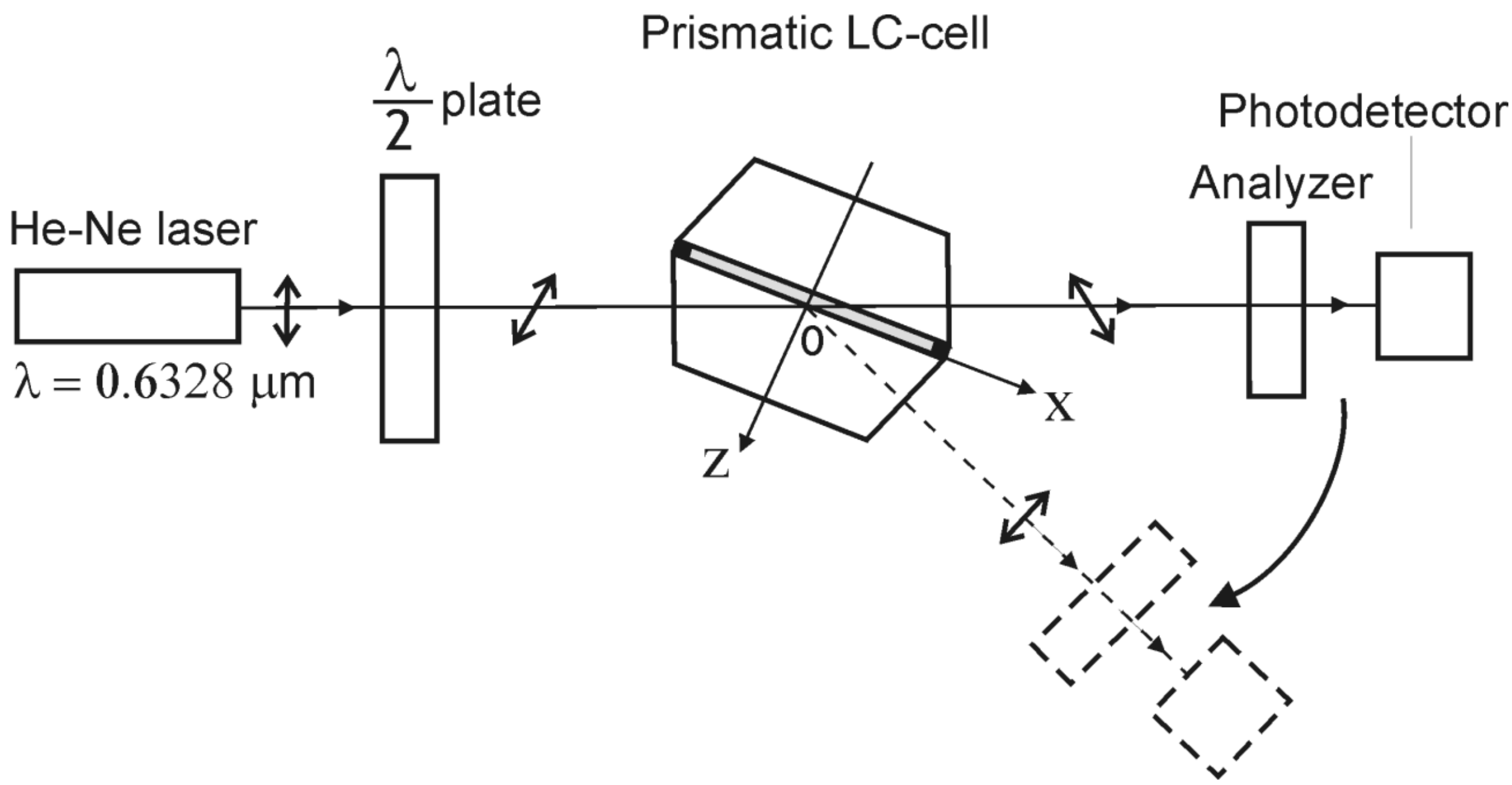

FIGURE 4. Experimental setup

The setup comprised a He-Ne laser emitting polarized radiation at a wavelength of 632.8 nm in a beam of 1- mm diameter. The laser beam passed through a half-wave plate and was directed onto an LC cell, which transmitted the extraordinary wave and reflected the ordinary wave due to the total internal reflection. The beam arrived at the prismatic cell normal to its entrance face. Consequently, the angle of light incidence on the LC layer, α, coincided with the angle at the base of prism.

The polarization azimuths of the passed extraordinary wave and reflected ordinary wave were measured using an analyzer and photodetector. To measure the polarization azimuth of the reflected ordinary wave, the half-wave plate 2 was rotated to a position at which only the ordinary wave was induced in the cell. The polarization azimuth of the passed extraordinary wave was measured at such position of the half-wave plate at which only this wave emerged in the cell.

The cell consisted of two trapezoidal glass prisms with the base 46×16 mm in size, base angle of 68°, and 16-mm height. The prisms were made of optical glass with a refractive index of 1.644 for a wavelength of 632.8 nm. The gap between the prisms was 25 μm. The cells were filled with nematic LC mixture №247 developed at Department of Chemistry, Vilnius University (Lithuania):

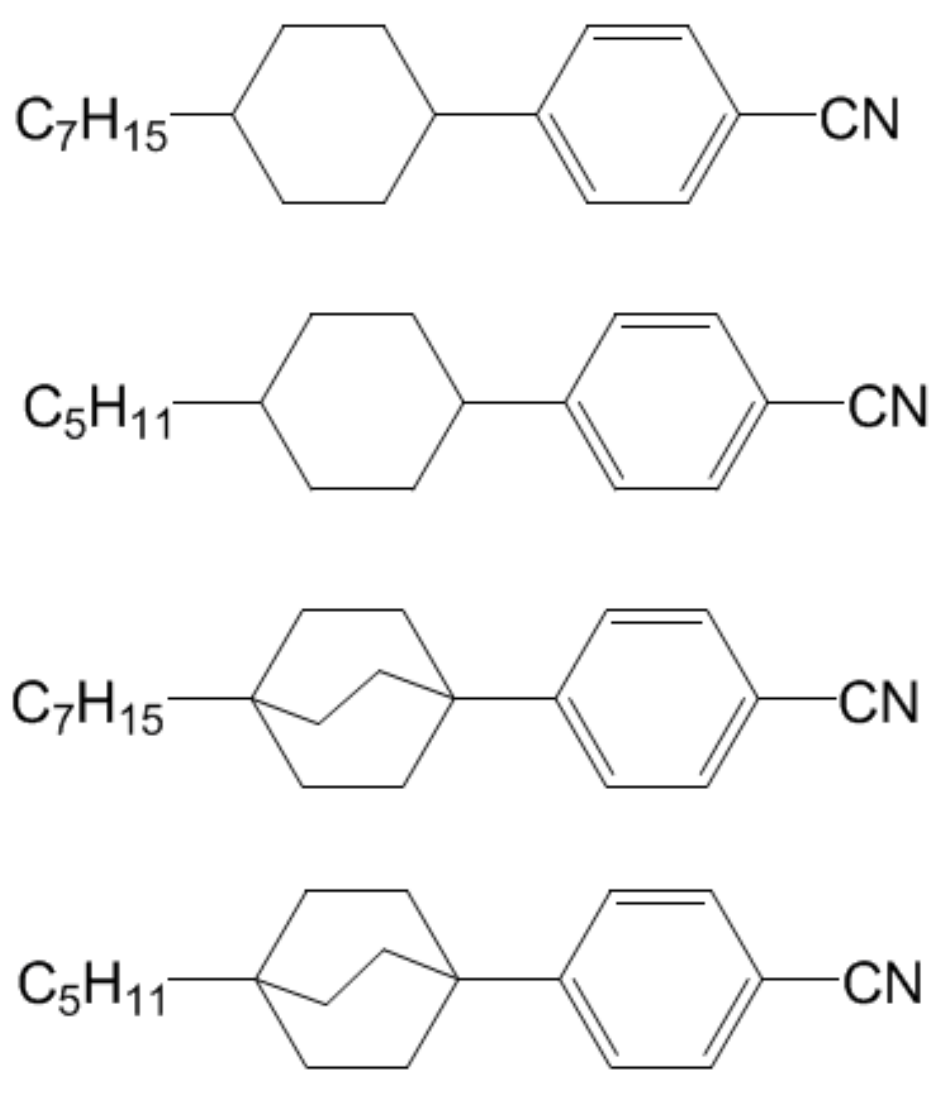

3:2:3:2 weight ratio.

The mixture is characterized by a clearing temperature of 76ºC and refractive indices $N_o = 1.492$ and $N_e = 1.610$ for $\lambda = 0.6328$ μm at a temperature of 20ºC. Conditions (1) and (2) are satisfied for the given refractive indices of the liquid crystal and glass and for the base angle of 68º of the prism. The deviations of the measured polarization azimuths of the reflected and passed waves from their true values, which were calculated for the given values of the angle of incidence and refractive indices of the liquid crystal and glass, did not exceed ±1º.

We made five LC cells. One cell with the homeotropic orientation was used to determine the zero azimuth of the analyzer (i. e., the position of the plane of incidence) from which angles $\beta_{e1}$, $\beta_{e2}$ and $\beta_{o1}$, $\beta_{o2}$ were counted. In the other cells, different hybrid types of orientation were created. To obtain the homeotropic orientation of the liquid crystal, we used a chromium stearyl chloride solution in isopropyl alcohol. The solution was applied to the bases of the glass prisms by spin coating at a speed of 3000 rpm. The applied layer was dried at a temperature of 120ºC for 40 min. The tilted orientation was created by rubbing this homeotropically aligning layer according to the known technique [16-17]. The layer was rubbed by a cotton cloth in the direction normal to the long axis of the base of the glass prism, that is, in the direction of the OY axis (see Fig. 1). The planar orientation was obtained by rubbing a polyvinyl alcohol layer (which was also applied by spin coating) in the same direction.

The orientation of the $\mathbf{n}_1$ and $\mathbf{n}_2$ directors at the LC layer interfaces in the cells I-IV is schematically shown in Fig. 5.

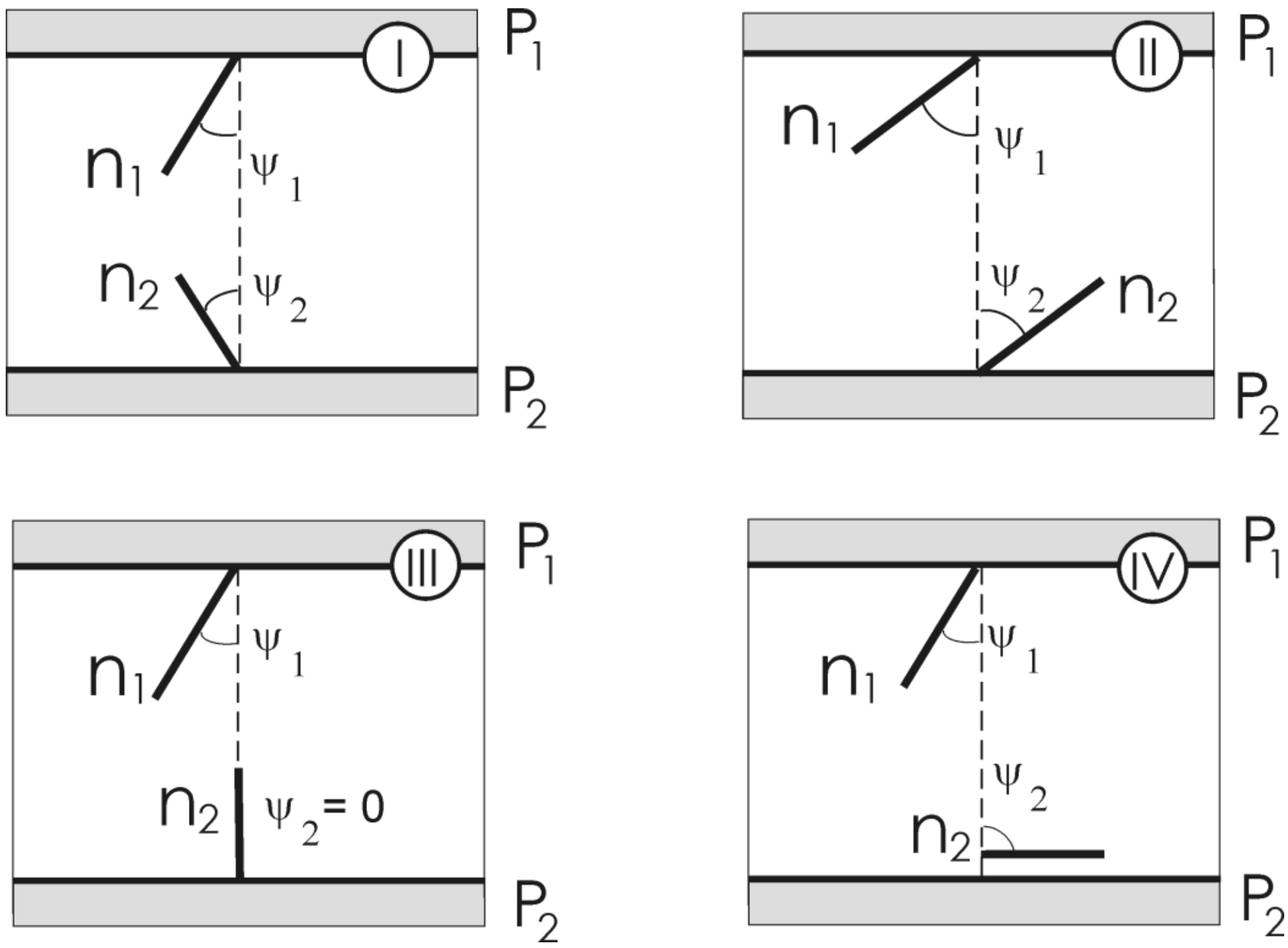

FIGURE 5. Orientation of the LC directors $\mathbf{n_1}$ and $\mathbf{n_2}$ at the $P_1$ and $P_2$ layer interfaces in cells I - IV.

To determine the director tilt angles at the interfaces of the LC layer, the measurements were made at two positions of the cell with respect to the direction of light incidence. In the first case, light was incident on the $P_1$ interface and the tilt angles thereat and at the $P_2$ interface were determined from the azimuth of the reflected ordinary wave $\beta_{o1}$, and the azimuth of the passed extraordinary wave $\beta_{e2}$, respectively. In the second case, light was directed onto the $P_2$ interface, and the tilt angles thereat and at the $P_1$ interface were determined from the azimuth of the reflected ordinary wave $\beta_{o2}$ and the azimuth of the passed extraordinary wave $\beta_{e1}$, respectively.

The values of each of the tilt angles measured in these two cases coincided within the accuracy of the experiment whose error we estimated at ±1º. The measurement results for the tilt angles $\Psi_1$ and $\Psi_2$ in four cells are listed in Table 1.

| *Cell* | $\Psi_1(°)$ | $\Psi_2(°)$ |
|---|---|---|
| I | 32 | 33 |
| II | 51 | 47 |
| III | 29 | 0 |
| IV | 25 | 89 |

**Table 1**. Director tilt angles $\Psi_1$ and $\Psi_2$ at the $P_1$ and $P_2$ interfaces

The tilt angles can be measured with much higher accuracy (for example, up to 0.1º), if high-precision polarimetric methods are used [18].

## Conclusion

We present a new method for measuring the director tilt angle at both interfaces of an LC layer. The method is based on measuring the polarization azimuth of the extraordinary wave passed through the LC cell or that of the ordinary wave reflected from the cell due to the total internal reflection. The polarization azimuth of light incident on the cell, which induces only the ordinary wave, can be used as well. The tilt angles of the LC director at both surfaces of an LC layer are related to the polarization azimuths of these waves through simple equations and can be readily calculated. The calculation does not call for the knowledge of the LC layer thickness and the refractive indices of glass and liquid crystal. The method is suitable for measuring the tilt angles within 0° to 90°. The proposed method was used for measuring the tilt angles of the director of a nematic liquid crystal in several LC cells.

## ACNOWLEGEMENTS

We are thankful to Dr. I.P. Kolomiets for his assistance in the method development. We also appreciate the use of the LC mixture kindly provided by Dr. P. Adomenas.